\documentclass[doublecol]{epl2} 

\usepackage{graphicx}
\usepackage[latin1]{inputenc}
\usepackage{graphicx,epstopdf}
\usepackage{dcolumn}
\usepackage{amsmath}
\usepackage[hypertex]{hyperref}
\usepackage{amsfonts}
\usepackage{amsthm}
\usepackage{amssymb}
\usepackage{mathrsfs}
\PassOptionsToPackage{caption=false}{subfig}
\usepackage{subfig}
\usepackage{color}
\usepackage[normalem]{ulem}

\title{Geometric classical and total correlations via trace distance}

\author{F. M. Paula\footnote{fagner@if.uff.br}, J. D. Montealegre\footnote{jdmontealegrer@if.uff.br}, A. Saguia\footnote{amen@if.uff.br}, Thiago R. de Oliveira\footnote{tro@if.uff.br}, \and M. S. Sarandy\footnote{msarandy@if.uff.br}}
\shortauthor{F. M. Paula \etal}

\institute{Instituto de F\'isica, Universidade Federal Fluminense, Av. Gal. Milton Tavares de Souza s/n, Gragoat\'a, 24210-346, Niter\'oi, RJ, Brazil}

\abstract{We introduce the concepts of geometric classical and total correlations through Schatten 1-norm (trace norm), which is the only Schatten $p$-norm able to ensure a well-defined geometric measure of correlations. In particular, we derive the analytical expressions for the case of two-qubit Bell-diagonal states, discussing the superadditivity of  
geometric correlations. As an illustration, we compare our results with the entropic correlations, discussing both their 
hierarchy and monotonicity properties. Moreover, we apply the geometric correlations to investigate the ground state of 
spin chains in the thermodynamic limit. In contrast to the entropic quantifiers, we show that the classical correlation  
is the only source of 1-norm geometric correlation that is able to signaling an infinite-order quantum phase transition.}

\begin{document}

\newcommand{\red}[1]{\textcolor{red}{#1}}
\newcommand{\blue}[1]{\textcolor{blue}{#1}}

\newcommand{\inprod}[2]{\braket{#1}{#2}}
\newcommand{\Tr}{\mathrm{Tr}}
\newcommand{\vp}{\vec{p}}
\newcommand{\Or}{\mathcal{O}}
\newcommand{\so}[1]{{\ignore{#1}}}

\maketitle

\section{Introduction}

Quantum discord has been introduced by Ollivier and Zurek~\cite{Ollivier} as an entropic measure of quantum 
correlation, whose characterization has attracted much attention during the last decade~\cite{Modi,Celeri,Sarandy:12,Gu:12}. 
From an analytical point of view, its evaluation is a difficult task, with closed expressions known only for 
certain classes of states~\cite{Luo,Ali,Giorda,Adesso,Li}. This difficulty led to the introduction of a geometric 
measure of quantum discord based on Schatten 2-norm (Hilbert-Schmidt metric) \cite{Dakic,Tufarelli:13}, which can be seen as a 
deformation of the entropic quantum discord \cite{Costa:13} computable for general bipartite states~\cite{Lu-Fu,Hassan,Rana}. 
Moreover, it has been shown to exhibit operational significance in specific quantum communication protocols~\cite{Dakic:12}. 
However, this Hilbert-Schmidt geometric discord cannot be regarded as a good measure for the quantumness of correlations, since it 
may increase under local trace-preserving quantum channels for the unmeasured party~\cite{Hu:12,Tufarelli:12,Piani}. 

Fortunately, this problem can be completely solved by adopting Schatten 1-norm (trace norm)~\cite{Hu:12,Nielsen-Chuang,Rana:13}. 
Indeed, concerning general Schatten $p$-norms for a geometric definition of quantum discord~\cite{Debarba}, it has been 
shown by some of the present authors in Ref.~\cite{Paula} that the 1-norm is the only $p$-norm able to consistently define a geometric quantum discord. In particular, for two-qubit systems, the geometric quantum discord based on Schatten 1-norm is equivalent to the negativity of 
quantumness~\cite{Nakano:12} (also referred to as the minimum entanglement potential \cite{Chaves:11}), which is a measure of nonclassicality 
recently introduced in Ref.~\cite{Piani:NoQ} and experimentally discussed in Ref.~\cite{Silva:12}. This quantifier has been calculated for general 
Bell-diagonal states~\cite{Paula,Nakano:12} and, more recently, has been extended for arbitrary X states~\cite{Ciccarello}. Thus, geometric 
quantum discord based on the 1-norm turns out to be even more computable than the entropic quantum discord. Moreover, it presents remarkable 
properties under decoherence for simple Bell-diagonal states as, for instance, double sudden change~\cite{Montealegre:13} and freezing 
behavior~\cite{Montealegre:13, Aaronson}.

In order to provide a unified view of correlations, a great deal of effort has been devoted to the investigation of the 
classical and total correlations in quantum-correlated systems~\cite{Groisman:05,Henderson:01,Bellomo:12,Bellomo:12-2,Modi:10}. 
For instance, the entropic quantum discord is defined by the difference between the quantum mutual information in a bipartite system 
before and after a local measurement is performed over one of the subsystems, with these two terms interpreted as the total and classical 
correlations, respectively~\cite{Groisman:05,Henderson:01}. In particular, these entropic quantifiers can be obtained as special cases of the 
generalized approach for correlations recently proposed by Brodutch and Modi~\cite{Brodutch}. 

In this work, we employ the approach by Brodutch 
and Modi to define geometric classical and total correlations via trace distance. Furthermore, we analytically evaluate these geometric correlations 
for the class of two-qubit Bell-diagonal states. As illustrations, we compare our results with the entropic quantifiers, analyzing additivity, hierarchy, 
and monotonicity relationships. Moreover, we also consider the geometric quantum correlations in the ground state of critical spin chains in the 
thermodynamic limit, discussing their behaviors at quantum phase transitions.
\section{Entropic and geometric correlations}

In the last decade, a number of measures for quantum correlations have been proposed. Remarkably, most of these measures can be described in a general framework proposed 
by Brodutch and Modi~\cite{Brodutch}. In this approach, one introduces a correlation measure using a general discord function $K[\rho,M(\rho)]$ defined by a distance (or pseudo-distance) 
between a state $\rho$ and the classical state $M(\rho)$, which results from a measurement on $\rho$ chosen according to pre-selected strategy. In terms of $ K[\rho,M(\rho)]$, 
the quantum, classical, and total correlations are then respectively given by
\begin{equation}
Q(\rho)=K[\rho, M(\rho)],\label{Q}
\end{equation}
\begin{equation}
C(\rho)=K[ M(\rho), M(\pi_{\rho})],\label{C}
\end{equation}
and
\begin{equation}
 T(\rho)=K[\rho,\pi_{\rho}],\label{T}
\end{equation}
where 
\begin{equation}
\pi_{\rho}=\rho_{1}\otimes \cdots \otimes \rho_{n}
\end{equation}
represents the product of the local marginals of $\rho$ in an $n$-partite system. One
of the possible strategies for the measurement $M$ is to choose it so as to minimize
the quantum correlation $Q$, which will be the approach adopted here. 
For qubit states, the minimization over projective measurements $M$ turns out to be 
equivalent to a minimization over all the set of classical states if we define the discord 
function $K[\rho,M(\rho)]$ in terms of relative entropy, Hilbert-Schmidt norm or trace 
distance~\cite{Paula,Nakano:12,Bellomo:12}. Given such a minimizing measurement $M$ for the 
quantum correlation $Q$, it will also be used in the evaluation of the classical correlation 
$C$, with no additional optimization required. An alternative approach for the classical and total correlations would be to consider an extra optimization to find out the closest product state \cite{Modi:10}. Even though this may lead to different results, it is equivalent to our framework in terms of formal criteria of correlation measures.

Here, we consider a two-qubit system, with parts labeled by $a$ and $b$, and we choose $M(\rho)$ as a classical-quantum state emerging from a projective measurement on 
subsystem $a$ that minimizes Eq.~(\ref{Q}), i.e.:
\begin{equation}
M(\rho)=\sum_{k=-,+}\left(\Pi_{k}\otimes \mathbb{I}\right)\rho \left(\Pi_{k}\otimes \mathbb{I}\right),
\end{equation}
where
\begin{equation}\label{PJ}
\Pi_{\pm}=\frac{1}{2}\left(\mathbb{I}\pm \vec{n}\cdot\vec{\sigma}\right)
\end{equation}
are projection operators, $\mathbb{I}$ is the identity matrix, $\vec{\sigma}=\left(\sigma_{1},\, \sigma_{2},\,\sigma_{3}\right)$ is a vector formed by Pauli matrices, and $\vec{n}=\left(n_{1},\, n_{2},\,n_{3}\right)$ is an unitary vector that minimizes $Q (\rho)$. 

In this scenario, the original entropic correlations ($Q_{E}$, $C_{E}$, $ T_{E}$) introduced by Olliver and Zurek~\cite{Ollivier} 
are obtained using a pseudo-distance function based on quantum mutual information $I(\rho)$~\cite{Brodutch}:
\begin{equation}
K_{E}[\rho, M(\rho)]=|I(\rho)-I(M(\rho))|,
\end{equation}
where $I\left( {\rho}\right) =S\left( {\rho}\parallel {\rho}_{A}\otimes {\rho}_{B}\right)$, 
with $S\left( {\rho}\parallel {\sigma}\right) =\text{tr}\left( {\rho} \log_2 {\rho}-{\rho}\log_2 {\sigma}\right)$ 
denoting the relative entropy. 

Moreover, by adopting a geometric point of view, we can also define geometric correlations ($Q_{G}$, $C_{G}$, $T_{G}$) in a 
consistent way through the trace distance~\cite{Paula,Nakano:12}:
\begin{equation}
K_{G}[\rho, M(\rho)]=\left\Vert \rho-M(\rho)\right\Vert_{1}=\text{tr}\left|\rho-M(\rho)\right|.
\end{equation}
The quantum part ($Q_{G}$) is known as the 1-norm geometric quantum discord (or negativity of quantumness). In contrast to the entropic quantum discord ($Q_{E}$), it can be analytically evaluated for an arbitrary two-qubit X state~\cite{Ciccarello}.
\section{Geometric classical and total correlations for Bell-diagonal states}

In order to compare geometric with entropic correlations, we will focus in the two-qubit Bell-diagonal states, i.e., a particular case of two-qubit X states whose density operator presents the form
\begin{equation}\label{bell}
\rho=\frac{1}{4}\left[\mathbb{I}\otimes \mathbb{I}+\vec{c}\cdot \left(\vec{\sigma}\otimes\vec{\sigma}\right)\right],
\end{equation}
where $\vec{c}=\left(c_{1},c_{2},c_{3}\right)$ is a three-dimensional vector composed by the correlation functions $c_{i}=\langle\sigma_{i}\otimes\sigma_{i}\rangle$. In this case, the expressions for the entropic correlations are well known \cite{Luo}, reading
\begin{equation}
Q_{E}=T_{E}-C_{E},
\end{equation}
\begin{equation}\label{CE}
C_{E}=\log_{2}[(1-c_{+})^{\frac{1-c_{+}}{2}}(1+c_{+})^{\frac{1+c_{+}}{2}}],
\end{equation} 
and
\begin{equation}
T_{E}=\log_{2}(4\lambda_{00}^{\lambda_{00}}\lambda_{01}^{\lambda_{01}}\lambda_{10}^{\lambda_{10}}\lambda_{11}^{\lambda_{11}}),
\end{equation} 
where 
\begin{equation}
c_{+}=\max\{|c_{1}|,|c_{2}|,|c_{3}|\}
\end{equation}
is the maximum among the elements of the set $\{|c_{1}|,|c_{2}|,|c_{3}|\}$ and
\begin{equation}
\lambda_{ij}=\frac{1}{4}\left[1+(-1)^{i}c_{1}-(-1)^{i+j}c_{2}+(-1)^{j}c_{3}\right]
\end{equation}
denotes the eigenvalues of the density operator $\rho$.

Concerning geometric correlations for Bell-diagonal states, an explicit expression has been obtained for $Q_{G}$ in Refs.~\cite{Paula, Nakano:12}. Here, we present an alternative derivation of $Q_{G}$ in order to identify the optimal classical-quantum state $M(\rho)$, which will be required for the definition of $C_{G}$ and $T_{G}$. Thus, let us start from the expression
\begin{equation}\label{QG1}
Q_{G}(\rho)=\text{tr}\left|\rho-M(\rho)\right|.
\end{equation}
The four possible eigenvalues of the operator $\rho-M(\rho)$ are given by  
$\gamma_{1}=\gamma_{+}$, $\gamma_{2}=-\gamma_{+}$, $\gamma_{3}=\gamma_{-}$ and $\gamma_{4}=-\gamma_{-}$, where 
\begin{equation}\label{GAMMA}
\gamma_{\pm}=\frac{1}{4}\sqrt{c^{2}-\vec{\alpha}\cdot\vec{u}\pm 2\sqrt{\vec{\beta}\cdot\vec{u}}}
\end{equation}
with $c^2=c_{1}^2+c_{2}^2+c_{3}^2$, $\vec{\alpha}=\left(c_{1}^2,c_{2}^2,c_{3}^2\right)$, $\vec{\beta}=\left(c_{2}^2c_{3}^2,c_{1}^2c_{3}^2,c_{1}^2c_{2}^2\right)$, and $\vec{u}=\left(n_{1}^2,n_{2}^2,n_{3}^2\right)$. Thus, Eq.~(\ref{QG1}) implies
\begin{equation}\label{QG2}
Q_{G}=\sum_{i=1}^{4}\left|\gamma_{i}\right|=2\left[\gamma_{-}(\vec{u})+\gamma_{+}(\vec{u})\right],
\end{equation}
where $\vec{u}$ minimizes the function $f(\vec{u})=\gamma_{-}(\vec{u})+\gamma_{+}(\vec{u})$ under the conditions $u_{1}+u_{2}+u_{3}=1$ and $0\leq u_{i}\leq 1$. Using the Lagrange multipliers method as described in Appendix~\ref{apa}, we conclude that $\vec{u}=\hat{\xi}_{j}$ ($j=1$, $2$, or $3$), where $\hat{\xi}_{j}$ represents the unitary vector in a fixed $c_{j}$ direction. Considering this result 
into Eq.~(\ref{QG2}), we find
\begin{equation}\label{QG3}
Q_{G} =\max\{|c_{j+1}|,|c_{j+2}|\},
\end{equation}
with the correlations $c_{j+1}$ and $c_{j+2}$ defined through modular arithmetics, i.e., 
$c_{j+k} = c_{j+k\,({\textrm{mod}} 3)}$. More specifically, we have $c_{2+2} = c_{3+1} = c_{1}$ and $c_{3+2} = c_{2}$. Defining the minimum
\begin{equation}
c_{-}=\min\{|c_{1}|,|c_{2}|,|c_{3}|\}
\end{equation}
and the intermediate 
\begin{equation}
c_{0}=\text{int}\{|c_{1}|,|c_{2}|,|c_{3}|\}
\end{equation}
values within the set $\{|c_{1}|,|c_{2}|,|c_{3}|\}$, it follows that Eq.~(\ref{QG3}) is minimized when $j$ is such that $|c_{j}|=c_{+}$, which corresponds to the following expression for quantum geometric correlation:
\begin{equation}\label{QG}
Q_{G} =\max\{c_{-},c_{0}\}=c_{0}.
\end{equation}

Now, substituting the unitary vector $\vec{n}=\pm\hat{\xi}_{j}$ that minimizes $Q_{G}$ in Eq.~(\ref{PJ}), we obtain the classical-quantum state   
\begin{equation}
M(\rho)=\frac{1}{4}\left(\mathbb{I}\otimes \mathbb{I}+c_{j}\sigma_{j}\otimes\sigma_{j}\right),
\end{equation}
which is still a Bell-diagonal state. Then, by computing the reduced density operators 
$\rho_{a}=\text{tr}_{b}\rho=\mathbb{I}/2$ and $\rho_{b}=\text{tr}_{a}\rho=\mathbb{I}/2$, 
we find a product state that is equal to the normalized $4\times4$ identity:
\begin{equation}
\pi_{\rho}=\rho_{a}\otimes\rho_{b}= \frac{1}{4}\left(\mathbb{I}\otimes \mathbb{I}\right).
\end{equation}
In this case, as $M(\pi_{\rho})=\pi_{\rho}$, Eqs.~(\ref{C}) and (\ref{T}) imply the following expressions for 
classical and total geometric correlations:
\begin{equation}
C_{G}=T_{G}(M(\rho))
\end{equation}
and
\begin{equation}
T_{G}=\text{tr}|\rho-\pi_{\rho}|=\sum_{i,j}\left|\lambda_{ij}-1/4\right|,
\end{equation}
where $\lambda_{ij}-1/4$ are the eigenvalues of operator $\rho-\pi_{\rho}$. In terms of $c_{+}$, $c_{0}$, and $c_{-}$, 
these expressions can be rewritten as
\begin{equation}\label{CG}
C_{G}=c_{+}
\end{equation}
and
\begin{equation}\label{TG}
T_{G}=\frac{1}{2}\left[c_{+}+\max\{c_{+},c_{0}+c_{-}\}\right].
\end{equation}
\section{Superadditivity, Hierarchy, and monotonicity properties}

\begin{figure}
\onefigure[scale=0.45]{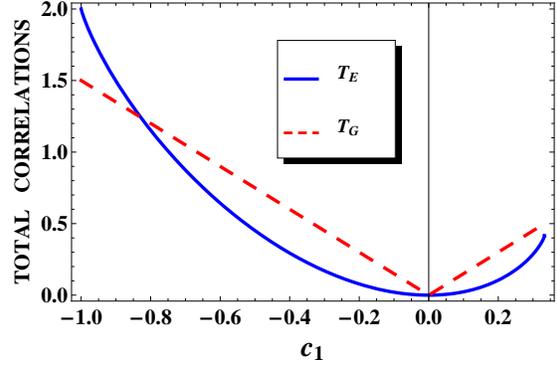}
\caption{(Color online) Plots of $T_{E}$ (solid line) and $T_{G}$ (dashed line) for 
SU(2)-symmetric states ($c_{1}=c_{2}=c_{3}$).}
\label{fig1}
\end{figure}

Let us now apply Eqs.~(\ref{QG}), (\ref{CG}) and, (\ref{TG}) to make general comparisons between $\left(Q_{G},C_{G},T_{G}\right)$ and $\left(Q_{E},C_{E},T_{E}\right)$. Firstly, we observe 
that the entropic correlations satisfy by definition the additivity relationship~\cite{Ollivier}, i.e.
\begin{equation}
T_{E}=C_{E}+Q_{E} .
\end{equation}
 On the other hand, Eqs.~(\ref{QG}),~(\ref{CG}), and ~(\ref{TG}) imply that the 1-norm geometric correlations satisfy a superadditivity relationship, i.e. 
\begin{equation}
T_{G}\leq C_{G}+Q_{G}\leq 2T_{G},
\end{equation}
with the equality $T_{G}=C_{G}+Q_{G}$ valid only for classical-quantum states, i.e., when $Q_{G}=0$. Regarding hierarchical relationships, we have
\begin{equation}\label{H1}
Q_{G}\geq Q_{E},
\end{equation}
\begin{equation}\label{H2}
C_{G}\geq C_{E},
\end{equation}
and
\begin{equation}\label{H3}
T_{G}\ngeqslant T_{E}.
\end{equation}
Eq.~(\ref{H1}) has been shown in Ref.~\cite{Paula},  Eq.~(\ref{H2}) can be deduced from Eq.~(\ref{CG}) by employing inequalities $c_{+}\geq\log_{2}(1-c_{+})(1+c_{+})$ and $1\geq(1\pm c_{+})/2$, whereas Eq.~(\ref{H3}) means that the hierarchy $T_{G}\geq T_{E}$ is not always true, as we can observed in Fig.~\ref{fig1}, which displays a cross between $T_{E}$ (solid curve) and $T_{G}$ (dashed curve) for SU(2)-symmetric states ($c_{1} = c_{2} = c_{3}$). Moreover, other hierarchies can be extracted by direct comparisons, yielding
\begin{equation}\label{H4}
T_{E}\geq C_{E},\,Q_{E},
\end{equation}
\begin{equation}\label{H5}
T_{G}\geq C_{G},\,Q_{G},
\end{equation}
\begin{equation}\label{H6}
C_{E}\ngeqslant Q_{E},
\end{equation}
\begin{equation}\label{H7}
C_{G}\geq  Q_{G}.
\end{equation}
The possibility $Q_{E}>C_{E}$ indicated by Eq.~(\ref{H6}) has been recently discussed in Ref.~\cite{Walczak:13}.

\begin{figure}
\onefigure[scale=0.45]{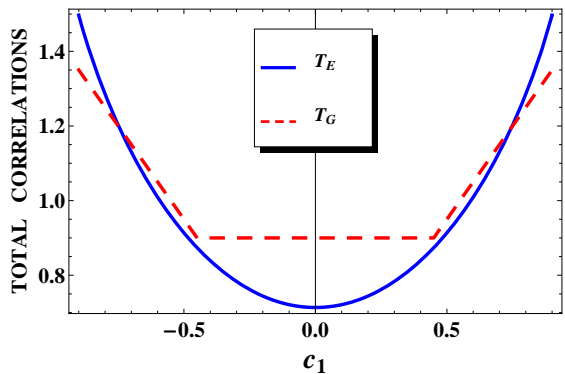}
\caption{(Color online) Plots of $T_{E}$ (solid line) and $T_{G}$ (dashed line) for 
U(1)-symmetric states ($c_{1}=c_{2}\neq c_{3}$) with $c_{3}=0.9$.}
\label{fig2}
\end{figure}

Concerning monotonicity relationships, it has been shown in Ref.~\cite{Paula} that the quantum correlations $Q_{G}$ and $Q_{E}$ maintain the ordering of states in the case of SU(2) symmetry, while such property is violated for more general classes of states, as in the case of U(1)-symmetric states (i.e., $c_{1}=c_{2}  \ne c_{3}$). This also occurs between the total correlations $T_{G}$ and $T_{E}$, as we can observe in Figs.~\ref{fig1} and \ref{fig2}. However, the classical correlations $C_{G}$ and $C_{E}$ are monotonic between themselves within the whole class of Bell-diagonal states. In fact, from Eqs.~(\ref{CE}) and (\ref{CG}), we find
\begin{equation}
C_{E}=\log_{2}[(1-C_{G})^{\frac{1-C_{G}}{2}}(1+C_{G})^{\frac{1+C_{G}}{2}}],
\end{equation}
where $C_{G}\in\left[0,1\right]$. Then, $C_{E}=0$ and $C_{E}=1$ when $C_{G}=0$ and $C_{G}=1$, respectively, and in the range $C_{G}\in\left(0,1\right)$, we have
\begin{equation}
\frac{dC_{E}}{dC_{G}}=\log_{2}\sqrt{\frac{1+C_{G}}{1-C_{G}}}> 0.
\end{equation}

It is important to emphasize that $T_{E}$ is a smooth function for arbitrary states \cite{Groisman:05}. On the other hand, $T_{G}$ may display sudden change behavior. As examples, Figs.~\ref{fig1} and \ref{fig2} illustrate sudden transitions of $T_{G}$ at points $c_{1}=0$ and $c_{1}=\pm 0.45$, respectively.
\section{Quantum spin chains}

\begin{figure}
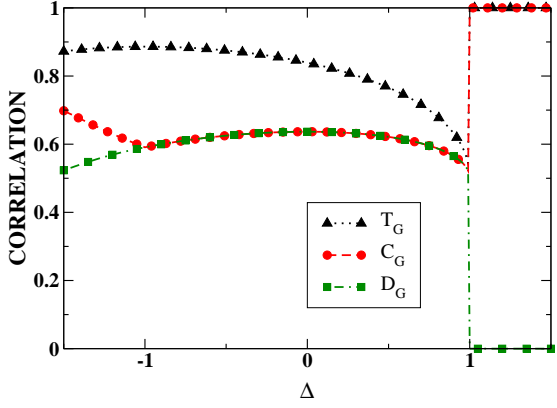

\onefigure[scale=0.3]{fig3.eps}
\caption{(Color online) Quantum ($Q_{G}$), classical ($C_{G}$), and total ($T_{G}$) geometric correlations as functions of the anisotropy $\Delta$ in the XXZ spin chain. Observe that $C_{G}$ is the only 1-norm 
geometric correlation able to characterize the infinite-order quantum phase transition ($\Delta=-1$).}
\label{fig3}
\end{figure}

Let us consider now applications in the critical properties of quantum spin chains whose ground states are described 
by Bell-diagonal states. In this context, we illustrate the discussion of the geometric correlations between two 
spins by investigating  the XXZ spin chain, whose Hamiltonian is given by
\begin{equation}
H_{XXZ}=-\frac{J}{2} \sum_{i=1}^{L} \left( \sigma^x_i \sigma^x_{i+1} + 
\sigma^y_i \sigma^y_{i+1} + \Delta \sigma^z_i \sigma^z_{i+1} \right),
\label{HXXZ}
\end{equation}
where periodic boundary conditions are assumed, ensuring therefore translation symmetry. 
We will set the energy scale such that $J=1$ and will be interested in a nearest-neighbor spin pair at 
sites $i$ and $i+1$. The model exhibits a first-order quantum phase transition at $\Delta=1$ and an 
infinite-order quantum phase transition at $\Delta=-1$. 
The entropic quantum discord has already been analyzed for nearest-neighbors, 
with the characterization of its quantum critical properties discussed~\cite{Sarandy:09,Dill:08}. 
Concerning its symmetries, the XXZ chain exhibits $U(1)$ invariance, namely, 
$\left[H,\sum_i \sigma_z^i\right]=0$, which ensures that the nearest-neighbor two-spin density operator is   
\begin{equation}
\mathcal{\rho}=\left( 
\begin{array}{cccc}
a & 0 & 0 & 0 \\ 
0 & b_1 & z & 0 \\ 
0 & z & b_2 & 0 \\ 
0& 0 & 0 & d
\end{array}
\right) .  \label{rhoAB}
\end{equation}
Moreover, the ground state has magnetization density $G^k_{z} = \langle \sigma_z^k \rangle=0$ ($\forall\, k$), 
which implies that 
\begin{eqnarray}
a &=& d = \frac{1}{4} \left(1+G_{zz}\right) \, , \nonumber \\
b_1 &=& b_2 = \frac{1}{4} \left(1-G_{zz}\right) \, , \nonumber \\
z &=& \frac{1}{4} \left(G_{xx}+G_{yy} \right).
\label{relem-xxz}
\end{eqnarray}
where, due to translation invariance, we have $G_{\alpha\beta}=\langle \sigma^1_\alpha \sigma^{2}_\beta \rangle=\langle \sigma^i_\alpha \sigma^{i+1}_\beta \rangle$ 
($\forall \, i$). By comparing the density operator given by Eq.~(\ref{rhoAB}) with the Bell-diagonal state given by Eq.~(\ref{bell}), 
we obtain that $c_1=c_2=2z$ and $c_3 = 4a-1$. By using the Hellmann-Feynman theorem~\cite{Hellmann:37,Feynman:39} for 
the XXZ  Hamiltonian~(\ref{HXXZ}), we then obtain
\begin{eqnarray}
c_1 &=& c_2 = \frac{1}{2} \left(G_{xx} + G_{yy}\right) = \Delta \frac{\partial \varepsilon_{xxz}}{\partial \Delta} 
- \varepsilon_{xxz} \, , \nonumber \\
c_3 &=& G_{zz} = -2 \frac{\partial \varepsilon_{xxz}}{\partial \Delta} \, ,
\label{c-xxz}
\end{eqnarray}
where $\varepsilon_{xxz}$ is the ground state energy density 
\begin{equation}
\varepsilon_{xxz} = \frac{\langle \psi_0| H_{XXZ} |\psi_0 \rangle}{L} = - \frac{1}{2} \left(G_{xx} + 
G_{yy} + \Delta G_{zz} \right),
\label{aux-xxz}
\end{equation}
with $|\psi_0\rangle$ denoting the ground state of $H_{XXZ}$. Eqs.~(\ref{c-xxz}) and~(\ref{aux-xxz}) hold for 
a chain with an arbitrary number of sites, allowing the discussion of correlations either for finite or infinite chains. 
Indeed, ground state energy as well as its derivatives can 
be exactly determined by the Bethe Ansatz technique for a chain in the thermodynamic limit~\cite{Yang:66}. The results 
obtained for geometric measures $Q_{G}$, $C_{G}$, and $T_{G}$ are plotted in Fig.~\ref{fig3}. 
We observe that all of the three measures are able to detect the first-order quantum phase transition at $\Delta = 1$, as in the case of entropic quantifiers \cite{Sarandy:09}. Concerning the infinite-order quantum phase transition at $\Delta = -1$, which occurs for $|G_{xx}|=|G_{zz}|$, the classical geometric correlation is given by $C_{G}=\max\left[\left|G_{xx}\right|,\left|G_{zz}\right|\right]$, consequently displaying a nonanalyticity in its derivative at the phase transition. On the other hand, the quantum and total correlations are given by $Q_{G} = \left|G_{xx}\right|$ and $T_{G} =|G_{xx}|+|G_{zz}|/2$ (around $\Delta \approx -1$). Therefore, these two measures do not change their behaviours at $\Delta = -1$, with no signature of this phase transition. Differently, the entropic correlation, $Q_{E}$, depends on $\max\left[\left|G_{xx}\right|,\left|G_{zz}\right|\right]$, signalling the transition at $\Delta = -1$ \cite{Sarandy:09}. Remarkably, it has recently been shown t
 hat this infinite-order critical point can be manifested in $Q_{G}$ as decoherence is taken into account~\cite{Montealegre:13}.
\section{Conclusions}

In summary, we have provided a unified view of geometric classical, quantum, and total correlations through trace distance. For the case of two-qubit Bell-diagonal states, we have found analytical expressions for such correlations. In particular, these expressions have allowed the investigation of a number of remarkable properties, such as the superadditivity of geometric correlations and both the hierarchy and monotonicity relations with the entropic measures. Moreover, we have applied our results to the ground state of critical spin chains, showing a special role played by classical correlations to characterize the infinite-order quantum phase transition. 

It is expected that the approach for unified geometric correlations introduced in this work may exhibit promising applications in a number of different scenarios, since it is
easily computable and constitutes a formally well-defined geometric measure. An an example, it can be applied to the geometric characterization of pointer 
basis and relaxation times in decohering systems. Moreover, it would be rather interesting to consider the extension of our results to the class of X states, 
which would allow to investigate geometric correlations in more general -- $Z_2$-symmetric -- critical systems, such as the transverse field XY model. These applications are left for future research.

\acknowledgments
This work is supported by the Brazilian agencies 
CNPq, CAPES, FAPERJ, and the Brazilian National Institute for Science and Technology of Quantum Information (INCT-IQ).

\section{Appendix}
\label{apa}
\renewcommand{\theequation}{A-\arabic{equation}}
\setcounter{equation}{0}

The aim of this Appendix is to show that the function defined by
\begin{equation}
f(u_{1},u_{2},u_{3})=\gamma_{-}(\vec{u})+\gamma_{+}(\vec{u}),
\end{equation}
where $\gamma_{-}(\vec{u})$ and $\gamma_{+}(\vec{u})$ are given by Eq.~(\ref{GAMMA}), under the constraint
\begin{equation}\label{eq:ge}
g(u_{1},u_{2},u_{3})=u_{1}+u_{2}+u_{3}=1,
\end{equation}
such that $u_{i}\in[0,1]$, always reaches its minimum for
\begin{equation}\label{eq:min}
\vec{u}=(1,0,0),\,\,(0,1,0),\,\,\,\text{or}\,\,\,(0,0,1).
\end{equation}

Let us start from the Langrange multipliers method, which consists in solving the system of equations
\begin{equation}\label{eq:sist}
\partial_{u_{l}}f=\lambda\partial_{u_{l}}g\,\,\,\,\,(l=1,\,2,\,3),
\end{equation}
where the variable $\lambda\in \Re$ is a Langrange multiplier. Note that $\partial_{u_{l}}g=1$, then $\lambda=\partial_{u_{l}}f$. By fixing $\lambda=\partial_{u_{j}}f$ ($j=1$, $2$, or $3$), the system of equations (\ref{eq:sist}) reduces to
\begin{equation}
\partial_{u_{m}}f-\partial_{u_{j}}f=0\,\,\,\,\,(m=\,j+1,\,j+2),
\end{equation}
with $u_{j+1}$ and $u_{j+2}$ defined through modular arithmetics, i.e., $u_{j+k} = u_{j+k\,({\textrm{mod}} 3)}$. After calculating the partial derivates, we obtain
\begin{equation}\nonumber
(\alpha_{m}-\alpha_{j})\left[\frac{1}{\gamma_{-}}\left(1-\frac{\alpha_{n}}{\sqrt{\vec{\beta}\cdot\vec{u}}}\right)+\frac{1}{\gamma_{+}}\left(1+\frac{\alpha_{n}}{\sqrt{\vec{\beta}\cdot\vec{u}}}\right)\right]=0
\end{equation}
\begin{equation}\label{eq:equ}
(n=\,j+1,\,j+2\,\,\text{and}\,\,n\neq m).
\end{equation}
When the equality (\ref{eq:equ}) is violated, it means that $f$ does not have an extremum in the open interval $u_{l}\in (0,1)$ and, therefore, the minimization must occur for one of the three vectors described in Eq.~(\ref{eq:min}). Thus, we only have to analyse situations where Eq.~(\ref{eq:equ}) is satisfied.

If $\alpha_{m}-\alpha_{j}\neq 0$, Eq.~(\ref{eq:equ}) can be satisfied only when $\gamma_{-}$, $\gamma_{+}$, and $\vec{\beta}\cdot\vec{u}$ are nonvanishing. In this case, we can divide Eq.~(\ref{eq:equ}) by $\alpha_{m}-\alpha_{j}$ and rewrite it as
\begin{equation}\label{eq:equ2}
\left(\alpha_{n}\vec{\alpha}+\vec{\beta}\right)\cdot\vec{u}+\alpha_{n}^{2}-c^{2}\alpha_{n}=0.
\end{equation}
Decomposing $\vec{u}$ in terms of the components $u_{m}$, $u_{n}$, and $u_{j}=1-u_{m}-u_{n}$ in Eq.~(\ref{eq:equ2}), we find $u_{n}=0$ and, consequently, $u_{j+1}=u_{j+2}=0$ and $u_{j}=1$. On the other hand, if at least two components of $\vec{\alpha}$ are equal (i.e., $\alpha_{j+1}=\alpha_{j+2}$), then
\begin{equation}\label{eq:func}
f=\left\{ 
\begin{array}{cc}
\frac{1}{2}\sqrt{\alpha_{j+1}}, & \alpha_{j+1}\geq \alpha_{j}  \\ \\
\frac{1}{2}\sqrt{\alpha_{j}+(\alpha_{j+1}-\alpha_{j})u_{j}}, & \alpha_{j+1}<\alpha_{j}
\end{array}
\right\}. 
\end{equation}
For $\alpha_{j+1}\geq \alpha_{j}$, any $\vec{u}$ minimizes $f$. When $\alpha_{j+1}<\alpha_{j}$, the minimization occurs for $u_{j}=1$ and $u_{j+1}=u_{j+2}=0$. This completes our proof.

\textit{Additional remark:} After completion of this work, we became aware of related results in Ref.~\cite{Aaronson:13}, where the approach introduced in Ref.~\cite{Modi:10} is adopted. Remarkably,  the geometric classical correlation obtained in Ref.~\cite{Aaronson:13} is monotonically related to the expression derived in our work.


\end{document}